\documentclass[]{raa}            
\usepackage{graphicx,times}
\usepackage{natbib}

\usepackage{amsmath}
\usepackage{amssymb}
\usepackage{multirow}
\usepackage{xcolor}

\providecommand{\bm}[1]{\mbox{\boldmath$#1$\unboldmath}}

\def\etal{et al.}

\begin{document}

   \title{Small glitches: the role of strange nuggets?
}

 \volnopage{ {\bf 2015} Vol.\ {\bf 9} No. {\bf XX}, 000--000}
   \setcounter{page}{1}

   \author{X.-Y. Lai
      \inst{1,2}
   \and R. -X. Xu
      \inst{3,4}
   }

   \institute{School of Physics, Xinjiang University, Urumqi 830046, China; {\it laixy@pku.edu.cn}\\
        \and
             Xinjiang Astronomical Observatory, Chinese Academy of Science, Urumqi 830011, China
        \and
             School of Physics, Peking University, Beijing 100871, China
         \and
             Kavli Institute for Astronomy and Astrophysics, Peking University, Beijing 100871, China\\
\vs \no
   {\small Received [year] [month] [day]; accepted [year] [month] [day] }
}

\abstract{
Pulsar glitches, i.e. the sudden spin-ups of pulsars, have been detected for most pulsars that we known.
The mechanism giving rise to this kind of phenomenon is uncertain, although a large data set has been built.
In the framework of star-quake model, based on~\cite{Baym1971}, the glitch-sizes (the relative increases of spin-frequencies during glitches) $\Delta \Omega/\Omega$ depend on the released energies during glitches, with less released energies corresponding to smaller glitch sizes.
On the other hand, as one of dark matter candidates, our Galaxy might be filled with the so called strange nuggets (SNs) which are the relics from the early Universe.
In this case the collisions between pulsars and SNs is inevitable, and these collisions would lead to glitches when enough elastic energy has been accumulated during the spin-down process.
The SNs-triggered glitches could release less energy, because the accumulated elastic energy would be less than that in the scenario of glitches without SNs.
Therefore, if a pulsar is hit frequently by SNs, it would tend to have more small size glitches, whose values of $\Delta \Omega/\Omega$ are smaller than that in standard star-quake model (with larger amounts of released energy).
Based on the assumption that in our Galaxy the distribution of SNs is similar to that of dark matter, as well as on the glitch data in ATNF Pulsar Catalogue and Jodrell Bank glitch table, we find that in our Galaxy the incidences of small size glitches exhibit tendencies consistent with the collision rates of pulsars and strange nuggets.
Further test of this scenario is expected by detecting more small glitches (e.g., by the Square Kilometre Array).
}

   \authorrunning{X.-Y. Lai \& R. -X. Xu}            
   \titlerunning{Small glitches: the role of strange nuggets?}  
   \maketitle

\section{Introduction}

Glitches are one type of pulsar timing irregularity, observed as discrete changes on the pulsar rotation rate.
Although a large data set for glitches has been built, the physical picture behind them still remains to be well understood.
Generally, there are two main models for the origin of glitches.
The first one is the star-quake model which regards a glitch as a star-quake of a spinning down pulsar resulting from the rearrangements of its crust~\citep{Baym1971}.
The second one regards glitches as the result of a rapid transfer of angular momentum between the inner superfluid and the outer crust~\citep{Anderson1975}.
Although the star-quake model is now less popular which falls to give a reasonable explanation for the glitch activity of Vela pulsar~\citep{Baym1971,Alpar1996}, the rearranges of the crust could be the trigger for the transfer of angular momentum described in the second model~\citep{Espinoza2011}.
What is certain to us is that, glitch behaviors reflect the interior structure of pulsars, and the data of glitches can give important information on the dense matter inside pulsars.
Some correlations between the glitch activities and the characteristic age of pulsars have been found, e.g., younger pulsars would have larger glitch spin-up rates $\dot{\Omega}$ and glitch-rates $\dot{N_g}$~\citep{Espinoza2011}.
The data of glitches make us to believe that glitches are caused by the changes of interior structures.
Nevertheless, can glitches related to some exterior trigger?

Another topic, seemly having no relation with glitch phenomenon, is about the relics of the early Universe.
In the Standard Model of cosmology, during the expanding history, our Universe underwent several kinds of phase transitions.
The QCD (quantum chromodynamics) phase transition, taking place when the age of the Universe was about 1 millisecond and the temperature was about 200 MeV, would leave out a huge amount of relics, if the phase transition was of the first order~\citep{Witten1984,Madsen1998}.
\cite{Witten1984} suggested that most of the deconfined quarks would be concentrated in the dense and invisible strange nuggets (SNs).
In fact, the formation and evolution of SNs depend on the physical properties of hot quark-gluon plasma and the state of dense matter, so the exact baryon number $A$ of each SN (or the mass-spectrum of SNs) is hard for us to determine.
Although some calculations were made, e.g.~\cite{Bhattacharjee1993,Bhattacharyya2000,Lugones2004}, the results were model dependent.
On the other hand, the properties of SNs could be constrained by astrophysics, if they do exist in the Universe.

As proposed by Witten, SNs could be a kind of dark matter candidate~\citep{Witten1984}.
Although being baryonic matter, SNs could behave like dark matter because of the very low charge-to-mass ratio, which comes from the fact that SNs are made of nearly equal numbers of $u$, $d$ and $s$ quarks.
As will be shown in this paper, the mass of each SN could be as high as $\sim 10^7$ g (with baryon number $A\sim 10^{34}$ and size $\sim 10^{-2}\ \rm cm$).
Whether such macroscopic and baryonic particles are compatible with observations, such as the formations of galaxies and stars, should be studied in some detail, but unfortunately even the numerical simulations in the framework of conventional microscopic dark matter particles have not reached consensus about the cosmic structure formation (e.g. see~\cite{Gao2007}).
A similar macroscopic dark matter candidate are the so called massive compact halo objects (MACHOs)~\citep{Paczynski1986}, and some constraints from observations have been found.
Very heavy MACHOs with masses above $10^{-7}M_\odot$ are excluded to be primary dark matter particles~\citep{EROS,Silk2007}, but dark matter could be made out of baryons and antibaryons in the form of compact stellar type objects or primordial black holes with sufficiently wide mass spectrum~\citep{Dolgov2009}.
The latest constrains on MACHOs could be found in~\cite{Monroy-Rodriguez}, which put the upper bound for the mass to be $5M_\odot$.
The studies about MACHOs could infer that, whether macroscopic dark matter particles with masses below $10^{-7}M_\odot$ have any conflict with either observations or simulations remains to be seen, and the related problem about how the SNs affect the structure formation is worth exploring in the future.

SNs as dark matter should not lead to conflicts with observations, and on the other hand the observations could give constraints on the properties of SNs.
\cite{Madsen1985} found that if dark matter is completely composed of SNs, the consistency between the prediction of Big Bang Nucleosynthesis (BBN) and the observational result for the helium abundance would not be affected if the radius of SNs exceed $\sim 10^{-5}$ cm, corresponding to the baryon number $A$ of each SN with $A\gtrsim 10^{24}$ (assuming that the matter inside SN has the density comparable to that of nuclear matter density).
SNs in the early Universe would affect the structure formation.
On the other hand, SNs could help us to understand some astrophysical phenomenons.
The formations of high-redshift ($z\sim 6$) supermassive black holes ($10^9M_{\odot}$)~\citep{SDSS2003} could be the results of SNs~\citep{Lai2010}, and in this picture the baryon number of each SN should be smaller than about $10^{35}$.
Another astrophysical consequence of SNs in our present Universe is that, if they pass our Earth, some seismic signals would be detected~\citep{seismic1,seismic2}.
More detailed studies about the effects of strange nuggets on some important processes, such as the formation of early galaxies and black holes, are expected.

In this paper, we propose another astrophysical consequence, which is related to the pulsar glitches.
If SNs formed in the early Universe and survived in the cosmic evolution, they would fill our Milky Way, and they would inevitably be accreted by stars.
With strong gravitational field, the accretion of SNs by pulsars would not be rare events.
In fact, the accretion of dark matter particles by sun as well as by pulsars have been studied~\citep{Press1985,Kouvaris2008}, and some interesting consequences have been discussed, i.e~\citep{Goldman1989,Huang2014}
Assuming that in our Galaxy the distribution of SNs is similar to that of dark matter, we can then know the accretion rates of SNs by pulsars.
Whether such accretion would affect the properties of glitches depends both on the states of matter inside pulsars and the glitch mechanism.
In this paper, we assume that pulsars are solid strange stars and apply the star-quake model~\citep{Baym1971} as the glitch mechanism.
In fact, the star-quake glitch models of solid strange stars have been studied~\citep{Zhou2004,Peng2008,Zhou2014}, which show that glitch behaviors and the energy released are consistent with observations.
Moreover, in star-quake model, the recovery process after a glitch could be explained as the damped oscillations~\citep{Zhou2004}.

In the framework of star-quake model~\citep{Baym1971}, the relation between glitch sizes (the relative increases of spin-frequencies during glitches) $\Delta \Omega/\Omega$ and the released energy during glitches can be derived straightforward, which shows that less amount of released energy during glitches would correspond to smaller glitch sizes.
Pulsar structure would be affected by the collisions between the pulsar and SNs, which could help the pulsar to release its elastic energy that accumulated during the spin-down process, probably manifesting as glitches detected.
In this scenario, the amount of energy released in this glitch $E_{\rm re}$ would be smaller than that in the scenario without SNs, as the energy released by the SN itself is negligible.
Therefore, for a particular pulsar, the incidence of small size glitches should be the same as its accretion rate of SNs.
Here the small size glitches are relatively small ones, whose values of $\Delta \Omega/\Omega$ are smaller than that in the standard star-quake scenario without SNs.

In this paper, we demonstrate the relation between accretion of SNs and the incidences of small size glitches in the star-quake model.
Based on the assumption that in our Galaxy the distribution of SNs is similar to that of dark matter, as well as on the glitch data in ATNF Pulsar Catalogue and Jodrell Bank glitch table, we find that in our Galaxy the incidences of small glitches exhibit a tendency consistent with the collision rates of pulsars and strange nuggets.
It should be noted that, although in this paper we assume pulsars are totally in solid state, our conclusions would not change quantitatively if pulsars only have solid crusts.

This paper is arranged as follows.
In \S 2 we will derive the relation between the glitch sizes and the released energies during a glitch, in the star-quake of solid strange star model.
Based on this, the relation between accretion of SNs and the small sized glitches will be demonstrated in \S 3.
In \S 4 we will get the expected accretion rates of SNs by pulsars and compare that with the observed event rates of small size glitches.
Conclusions and discussions are made in \S 5.

\section{Glitches in Star-quake Model}

Following~\cite{Baym1971}, we will derive the released energies in terms of the glitch sizes (the relative increases of spin-frequencies during glitches) $\Delta \Omega/\Omega$.
In addition, we do the calculation in the framework of solid strange star model, which means that all parts of the stars are in solid state, but the results would not change quantitatively in the model of neutron stars with solid crusts.
When a solid star spins down, strain energy develops until the stresses reach a critical value.
At that moment, a star-quake occurs and the stresses are relieved and its rotation rate is suddenly increased by conservation of angular momentum (similar calculations can be found in~\cite{Zhou2004} and~\cite{Zhou2014}).

Total energy of a rotating star with mass $M$, radius $R$ and angular momentum $L$ is
\begin{equation}
E=E_0+\frac{L^2}{2I}+A\epsilon^2+B(\epsilon-\epsilon_0)^2,
\end{equation}
where $E_0$ is the total energy in the non-rotating case, and
\begin{equation}
\epsilon=\frac{I-I_0}{I_0}
\end{equation}
is the oblateness of the star with moment of inertia $I$, which reduces to $I_0=0.4MR^2$ in the non-rotating case, and $\epsilon_0$ is the initial oblateness.
The coefficient measure the departure of gravitational energy relative to the non-rotating case is
$A=\frac{3}{25}GM^2/R$,
and the coefficient measure the strain energy is
$B=\frac{57}{50}VC_{44}$
where $C_{44}$ is the shear modulus and $V=4\pi R^3/3$ is the whole volume of the star~\citep{Baym1971}.
The angular momentum is conserved during a glitch, so the sudden chance in oblateness $\Delta \epsilon$
is directly related to the relative jump in angular momentum $\Omega$
\begin{equation}
\Delta \epsilon=\epsilon_{-i}-\epsilon_{+i}=-\frac{\Delta \Omega_{i}}{\Omega_{+i}}=\frac{\Omega_{+i}-\Omega_{-i}}{\Omega_{+i}},
\end{equation}
where the spin-up rate mentioned above $\Delta\Omega/\Omega$ is just $\Delta \Omega_i/\Omega_{+i}$ during $i$-th glitch, and the subscripts ``$+i$'' and ``$-i$'' denote quantities right before and right after $i$-th glitch.
The initial oblateness before $i$-th glitch $\epsilon_{0i}$ is~\citep{Baym1971}
\begin{equation}
\epsilon_{+i}=\frac{I_0\Omega_{+i}^2}{4(A+B)}+\frac{B}{A+B}\epsilon_{0i}.
\end{equation}
The evolution of $\epsilon$ as time is shown in~\cite{Zhou2004}.

For simplicity we assume that when a star-quake occurs, the entire stresses are relieved and the released energy is $E_{\rm re}$.
The energy conservation gives
\begin{equation}
\frac{1}{2}I_{+i}\Omega_{+i}^2+A\epsilon_{+i}^2+B(\epsilon_{+i}-\epsilon_{0i})^2=\frac{1}{2}I_{-i}\Omega_{-i}^2+A\epsilon_{-i}^2+E_{\rm re},\label{energy}
\end{equation}
then it is easy to get
\begin{equation}
\frac{\Delta{\Omega_i}}{\Omega_{+i}}=\frac{E_{\rm re}\mu}{2B\sigma_c}=\frac{2E_{\rm re}(A+B)}{BI_0}\frac{1}{\Omega_{0i}^2-\Omega_{+i}^2}.
\end{equation}
where $\sigma_c$ is the critical mean stress of the star, which corresponds to the occurrence of a glitch, and $\mu=2B/V$  is the mean shear modulus of the star.

The value of parameter $B$ for solid strange stars is uncertain now, because the structure and state of quark matter inside pulsars are still unsolved problems.
The shear modulus $C_{44}$ for a bcc lattice of nuclei of number density $n$, charge $Ze$ and lattice constant $a$, interacting via an unscreened Coulomb interaction, could be written as~\citep{Baym1971} $C_{44}\sim 0.4 Z^2e^2n/a\sim 10^{30}\ \rm erg/cm^3$.
Based on the fact that the strong interaction is about 2 to 3 orders of magnitude stronger than electromagnetic interaction, we infer that the value of $C_{44}$ of solid strange stars is about in the range $10^{32}-10^{33}\ \rm erg/cm^3$~\citep{Xu2003}, then $B=\frac{57}{50}VC_{44}\sim10^{52}\ {\rm erg}\cdot (C_{44}/10^{33}\ {\rm erg/cm^3})(R/10{\rm\ km})^3$.
Comparing with $A\sim 6\times 10^{52}\ {\rm erg}\ (M/1.4\ M_{\odot})^2\cdot(10\ {\rm km}/R)$, we can approximate that $(A+B)/B\simeq 10$, then
\begin{eqnarray}
\frac{\Delta{\Omega_i}}{\Omega_{+i}}&\simeq&\frac{E_{\rm re}(A+B)}{BI_0\Omega_{0i}}\frac{1}{|\dot{\Omega}|\Delta t}\nonumber\\&\simeq&10^{-6}\left(\frac{E_{\rm re}}{10^{34}\ \rm erg}\right)\left(\frac{10^{45}\ \rm g\cdot cm^2}{I_0}\right)\left(\frac{P}{1\ \rm s}\right)^3\left(\frac{10^{-15}}{\dot{P}}\right)\left(\frac{10^7\ \rm s}{\Delta t}\right),\label{gs}
\end{eqnarray}
where $P=2\pi/\Omega$ is the spin period, and $\Delta t$ is the time interval between two successive glitches.
The result is consistent with Fig.2 in~\cite{Zhou2014} which shows the upper limit of the released energy during glitches (for bulk invariable case).

\section{strange nuggets and Glitches}
\label{SNs and glitches}

\subsection{Cosmological QCD phase transition and strange nuggets (SNs)}

Cosmological QCD (quantum chromodynamics) phase transition, taking place when the age of the Universe was about $10^{-6}$ seconds and the temperature was about 200 MeV, could have interesting astrophysical consequences.
If it was of the first order, a huge amount of strange nuggets (SNs) would form and survive the cooling of the Universe, and SNs were proposed as dark matter candidates~\citep{Witten1984} although they are baryonic, due to their extremely low charge-to-mass ratio (a detailed review of SNs could be found in~\cite{Madsen1998}).
SNs are composed of up, down and strange quarks, with densities higher than nuclear matter density.
The state of SNs is uncertain, because this problem is related to the low-energy QCD theories, the same problem about the interior structure of pulsar-like compact stars.
SNs with small baryon number $A$ would have been evaporated in the early Universe, but the lower limit for the baryon number $A$ of the stable SNs is difficult for us to derive.
On the other hand, some constraints on size of SNs have been made combined with their astrophysical consequences.
If the existence of SNs would not break the consistency between the prediction of Big Bang Nucleosynthesis and observed result for the helium abundance, the baryon number of each SN should be larger than about $10^{24}$ (assuming that SNs have a uniform size)~\citep{Madsen1985}.
Using some more accurate data, the lower limit of baryon number has been improved to be $\sim 10^{25}$~\citep{Lai2010}.
If the formations of high-redshift ($z\sim 6$) supermassive black holes ($10^9M_{\odot}$)~\citep{SDSS2003} were the results of SNs, then the baryon number of each SN should be smaller than about $10^{35}$~\citep{Lai2010}.

Then what could SNs do in our present Galaxy?
As shown in~\cite{Witten1984}, the mass fraction of SNs could be as large as that of dark matter.
If SNs are stable and survive in present Universe, they would fill our Galaxy, with the total mass comparable with that of dark matter.
Circulating in the Galaxy, they would be accreted by stellar objects, especially pulsars whose gravitational fields are strong, and the accretion rate by a particular pulsar is then related to its position in the Galaxy.

Then we will derive the accretion rates of SNs by pulsars.
Assuming that SNs have the same distribution as that of dark matter, and the fraction of total mass of SNs and compared to the total mass of dark matter is $\eta$, whose value is chosen to be 0.1 in the following calculations.
For the dark matter density distribution and the function of distance from Galaxy center $r$, we choose the NFW form~\citep{NFW}
\begin{equation}
\rho_{\rm dm}(r)=\frac{\rho_s}{\left(\frac{r}{r_s}\right)\left[1+\left(\frac{r}{r_s}\right)^2\right]},
\end{equation}
where $\rho_s=0.26 \rm GeV\ cm^{-3}$ and $r_s=20$ kpc.
The number density of SNs is $n_x=\eta\cdot\rho_{\rm dm}/m_x$, where $m_x$ is the mass of each SN.
For simplicity, we assume that all of SNs have the same mass, with $m_x=A\cdot m_b$, where $A$ is the baryon number of each SN and $m_b\sim 1$ GeV is the mass of baryon.
The circular velocity $v$ of each SN is then
\begin{equation}
v(r)=\sqrt{\frac{GM(r)}{r}}
\end{equation}
where $M(r)=\int_0^{r}\rho(r^{\prime})4\pi r^{\prime2}dr^{\prime}$ is the total mass inside the sphere with radius $r$.
To derive the accretion rates of SNs by pulsars, we can borrow the results for the accretion rates of dark matter particles by stars/pulsars.
The accretion rate of SNs with number density $n_x$ by a pulsar with mass $M$ and radius $R$ is~\citep{Press1985,Kouvaris2008}
\begin{equation}
\mathcal{F}=4\pi^2
n_x\left(\frac{3}{2\pi v^2}\right)^{3/2}\frac{2GMR}{1-2GM/R}\
\frac{1}{3}v^2, \label{flux}
\end{equation}
where the factor $(1-2GM/R)^{-1}$ is the modification factor taking into account the general relativity effect.

\subsection{SNs and glitches}

The sample of all of the observed glitches also shows that, the glitch sizes $\Delta \Omega/\Omega$ spread between $10^{-11}$ and $10^{-4}$ with a bimodal distribution which peaks at approximately $10^{-9}$ and $10^{-6}$, respectively~\citep{Espinoza2011,Yu2013}.
The glitch sizes have not been well explained, but Eq.(\ref{gs}) tells us that the glitch sizes are related to the released energies during glitches.
When the stresses reach a critical value, the star releases strain energies $E_{\rm re}$, which are stored in the star during the spinning down process due to the rigidity of the star.
The released energy $E_{\rm re}$ in a glitch can be inferred from observations, if it is large enough for us to detect.
For example, a constraint on the X-ray flux enhancement of Vela 35 days after a glitch with $\Delta \Omega/\Omega\sim 10^{-6}$ have been made~\citep{Helfand2001}, which infer that the upper limit of the energy released is about $10^{36}$ erg.
If a pulsar accretes an SN, its structure would be affected, as the SN would interact with particles inside the pulsar.
The detail of such interaction is hard to derive, because it is related to both the interior states of pulsars and SNs.
For an estimation we take the parameters used by~\cite{Zhou2004}, where glitches occur as the consequence of the star-quake of solid strange stars.
When the stress reaches a critical value $\sigma_c$, $(\epsilon-\epsilon_0)\sim 10^{-3}$, then the strain energy density $\rho_{\rm strain}$ could be approximated as
\begin{eqnarray}
\rho_{\rm strain}&\simeq&C_{44}\cdot (\epsilon-\epsilon_0)^2\nonumber\\&\simeq&10^{-7}\ {\rm MeV\cdot fm^{-3}}\left(\frac{C_{44}}{10^{32}\ {\rm erg\cdot cm^{-3}}}\right)\left(\frac{\epsilon-\epsilon_0}{10^{-3}}\right)^2.
\end{eqnarray}
If the lattice (quark or quark-cluster) density inside pulsar is $\sim 0.3\ \rm fm^{-3}$, then to break the lattice structure each lattice should obtain about $10^{-6}$ MeV.
Just before hitting the surface of a pulsar, the energy of an SN is about $(A/10^{33})\times 10^{35}$ MeV, so if each lattice it goes through gains energy $V$, then the total number of lattices it would interact with is $N\sim 10^{41}\cdot(A/10^{33})\cdot (10^{-6}\ {\rm MeV}/V)$.
The cross section of the interaction between an SN and the particles inside the pulsar could be approximated as the geometric cross section of the SN, which is $\sim 10^{-4}\ {\rm cm^2}\cdot (A/10^{33})^{2/3}$, so the length of trajectory is $\sim 10\ {\rm km}\cdot (A/10^{33})^{1/3}\cdot (10^{-6}\ {\rm MeV}/V)$.

Although the exact demonstration about the breaking criterion is unknown, the above estimation shows that it is possible for an SN with $A\gtrsim 10^{33}$ to break the whole solid body (or crust) of the star, which would give rise to a glitch.
The amount of energy releasing of such glitches triggered by SNs would be smaller than that of the normal glitches, as the latter occur when the elastic energy accumulated reaches a critical value.
According to Eq(\ref{gs}), in the star-quake model the glitch size $\Delta\Omega/\Omega$ is proportional to the amount of energy releasing $E_{\rm re}$, so the sizes of SN-triggered glitches should be smaller than that of the normal glitches.

The amounts of energy releasing $E_{\rm re}$ in normal glitches are also difficult for us to quantify.
It should be different from pulsar to pulsar, and could even be different between different glitch-events of one pulsar.
The released energy is mainly from the accumulated elastic energy, and it could be inferred that the more massive pulsars should release more energy during glitches.
This dependence of $E_{\rm re}$ on the pulsar mass $M$ could eliminate some uncertainties of both $E_{\rm re}$ and the moment of inertia $I_0$.
Consequently, from Eq(\ref{gs}) we can infer that, we could use typical values of both $E_{\rm re}$ and $I_0$ to reduce the difference of $\Delta\Omega/\Omega$ between different pulsars.
Because the upper limit of the energy released in one glitch event of Vela (PSR B0833-45) pulsar is about $10^{36}$ erg~\citep{Helfand2001,Zhou2014}, it seems reasonable to set $E_{\rm re}=10^{35}$ erg for normal glitches.

Based on the above arguments, we could estimate the occur rates of SN-triggered glitches, which may be identified as those with $\Delta\Omega/\Omega<\Delta\Omega/\Omega(E_{\rm re}=10^{35}\rm erg)$.
On the other hand, we take Eq.(\ref{flux}) has the expected occur rates for SN-triggered glitches.
Combined with the observational data, we could then compare these two rates for different pulsars.

It is worth mentioning that, although the accretions of SNs by pulsars would lead to bursts in X-rays, it is not easy for them to be detectable.
The gravitational energy released by accretion of an SN with $A=10^{34}$ on a typical pulsar is about $10^{30}$ erg.
With degenerate electrons with Fermi energy $\sim 10$ MeV, the heat conductivity of strange matter must be very high, which means that strange stars will be thermalized by any heat flow.
The process of thermalization is complicated, but the time-scale $\tau_{\rm enh}$ for releasing enhanced X-ray bursts resulting from the accretions could be estimated as $\tau_{\rm enh} \gg \tau \sim 10^{30}\ {\rm erg}/L_{\rm bol}\sim 0.1\ {\rm s}\cdot (10^{31}\ {\rm erg\ s^{-1}}/L_{\rm bol})$, where $L_{\rm bol}$ is the bolometric X-ray luminosity (see~\cite{Yu2011} and references therein).
The luminosity of an enhanced X-ray burst resulting from one accretion is then $L_{\rm enh}=10^{30}\ {\rm erg}/\tau_{\rm enh}\ll 10^{30}\ \rm erg/\tau\sim 10^{31}{\rm erg\ s^{-1}}$, i.e., $L_{\rm enh}$ is much smaller than the bolometric X-ray luminosity.
Therefore, the accretions of SNs by some near and well-monitored compact stars, such as RX J0720.4-3125 and RX J1856-3754, would lead to X-ray variability which are difficult to be detected, so the accretions would not bring conflict with observations.

\section{Comparison of accretion rates of SNs by pulsars and glitch data}

As demonstrated in \S~\ref{SNs and glitches}, the collision between an SN and the pulsar could result to a small size glitches compared to that in the scenario without SNs.
To compare with the glitch data, we choose 25 pulsars whose recorded numbers of glitches are larger than 5~\citep{Espinoza2011}.
The data for glitch sizes $\Delta\Omega/\Omega$ are from Jodrell Bank glitch-tables~\citep{Espinoza2011} (http://www.jb.man.ac.uk/pulsar/glitches/html), and the rotation periods and the time derivation of periods, as well as the locations of pulsars compared to the barycentre of the solar system (which are used to derive the distances of pulsars to the Milky Way center) are from ATNF Pulsar Catalogue~\citep{ATNF} (http://www.atnf.csiro.au/people/pulsar/psrcat/).

The accretion rates of SNs by a pulsar (which are taken to be the expected event-rates of small glitches) with mass $M=1.4M_\odot$ and radius $R=10$ km and located at a distance $r$ from the Galaxy center are shown in Fig.1, where solid, dashed and dash-dotted lines correspond to three values of baryon number $A$ of each SN respectively.
These lines denote the expected occur rates for small glitches with $\Delta\Omega/\Omega<\Delta\Omega/\Omega(E_{\rm re}=10^{35}\rm erg)$.
The data points (blue stars) show the observed event-rates of small size glitches with $\Delta\Omega/\Omega<\Delta\Omega/\Omega(E_{\rm re}=10^{35}\rm erg)$ and the distances from the Milky Way center of the corresponding pulsars.
Here the event-rates are derived as the ratio of the number of glitches satisfying $\Delta\Omega/\Omega<\Delta\Omega/\Omega(E_{\rm re}=10^{35}\rm erg)$ to the whole time interval from the first and the last observed glitches.
The axes in Fig.1 are scaled in logarithm, so the points corresponding to zero observed value are not shown in the figure, expect for J0537-6910.
The observed value of J0537-6910 is shown on the horizontal axis although the value is zero, because it is the most distant one from the MW center among all these pulsars, and is consistent with our expectation.
In the calculations, we assume that SNs constitute 10\% of dark matter ($\eta=0.1$).
Although the mass ratio of SNs to dark matter $\eta$ is unknown, it is proportional to $A$ under a certain accretion rate.
For instance, if $\eta=0.01$, then the values of $A$ should be one tenth of that in Fig.1.

   \begin{figure}[h!!!]
   \centering
   \includegraphics[width=9.0cm, angle=0]{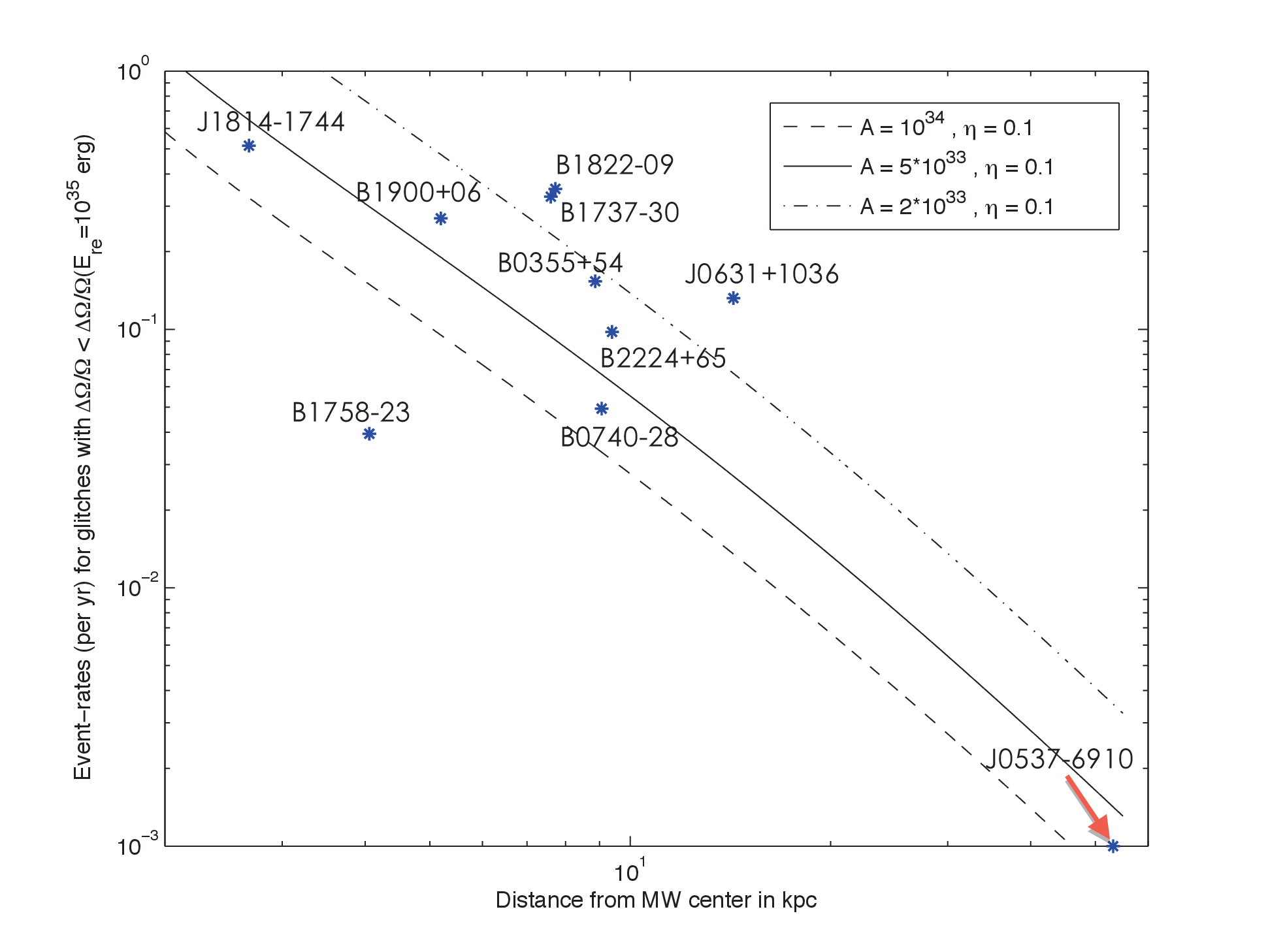}

   \begin{minipage}[]{85mm}

   \caption{The accretion rate of SNs, with baryon number $A=10^{34}$, $A=5\times 10^{33}$ and $A=2\times 10^{33}$, by a pulsar located in the distance of $r$ from the center of the Milky Way, are plotted in dashed, solid and dash-dotted lines respectively.
   In this paper, these lines denote the expect occur rates for pulsars with $\Delta\Omega/\Omega<\Delta\Omega/\Omega(E_{\rm re}=10^{35}\rm erg)$.
Data points: Observed rate per year of glitch-events with $\Delta\Omega/\Omega<\Delta\Omega/\Omega(E_{\rm re}=10^{35}\rm erg)$, for pulsars whose recorded glitch-number are larger than 5.
The data point corresponding to J0537-6910 in the horizontal axis corresponds to zero observed value.} \end{minipage}
   \label{fig2}
   \end{figure}

We can see that, except for the pulsars with zero event rate, the event-rates for glitches with $\Delta\Omega/\Omega<\Delta\Omega/\Omega(E_{\rm re}=10^{35}\rm erg)$ decrease with distances of pulsars from the Milky Way center, which show the same tendency as the accretion rates of SNs by pulsars.
To demonstrate the correlation of glitch rates on the distances from the Galactic center, we use the Pearson correlation coefficient $\rho$, which is a measure of the linear dependence between two variables~\footnote{For two variables $x$ and $y$, $\rho$ is defined as $\rho=\frac{\sum_i (x_i-\bar x)(y_i-\bar y)}{\sqrt{\sum_i (x_i-\bar x)^2}\sqrt{\sum_i (y_i-\bar y)^2}}$.
In general $|\rho|\leqslant 1$, where $\rho=1$ represents total positive correlation, $\rho=0$ represents no correlation, and $\rho=-1$ represents total negative correlation.}.
The result for data in Fig.1 (exclusive of J0537-6910) is $\rho\simeq -0.5$.
Although the interpretation of a certain value of $\rho$ will be different for different context and purposes without a criterion for all problems, the value $-0.5$ may imply the trend that, the more distant of a pulsar to the Galactic center, the less rates of small glitches it will have.

The dispersion of data points might mainly result from two factors: (1) the fact that not all the SNs have the same baryon number $A$; and (2) the mass and radius of each pulsar would be different from the typical values.
All of the 25 pulsars are listed in Table 1, including their spin-down ages, numbers of glitches (NGlt) and distances from the Milky Way center.
The upper part of the table includes 9 pulsars who have glitch events with $\Delta\Omega/\Omega<\Delta\Omega/\Omega(E_{\rm re}=10^{35}\rm erg)$, and the lower part of the table includes 16 pulsars who have no glitch event with $\Delta\Omega/\Omega<\Delta\Omega/\Omega(E_{\rm re}=10^{35}\rm erg)$.

\begin{table}[h!!!]

\small
\centering

\begin{minipage}[]{120mm}

\caption[]{25 Pulsars Whose Recorded Glitch Numbers (NGlt) Are Larger Than 5}\label{Table 1}\end{minipage}
\tabcolsep 6mm
 \begin{tabular}{clclcl}
  \hline\noalign{\smallskip}
Pulsar name &  Spin-down age$^1$ & NGlt$^2$ & Distance from the \\ & \ \ \ \ \ \ \ \ (yr) & & MW center$^1$ (kpc)                  \\
  \hline\noalign{\smallskip}
B0355+54  & $5.64\times 10^{5}$ & 6 & 8.87    \\
J0631+1036 &  $4.36\times 10^4$ & 15 & 14.29 \\
B0740-28 & $1.57\times 10^5$ & 8 & 9.06 \\
B1737-30 & $2.06\times 10^4$ & 33 & 7.60 \\
B1758-23 & $5.83\times 10^4$ & 12 & 4.06 \\
J1814-1744 & $8.46\times 10^4$ & 7 & 2.67 \\
B1822-09 & $2.32\times 10^5$ & 6 & 7.72 \\
B1900+06 & $1.38\times 10^6$ & 6 & 5.19 \\
B2224+65 & $1.12\times 10^6$ & 5 & 9.39 \\
\hline\noalign{\smallskip}
J0205+6449 & $5.37\times 10^3$ & 6 & 10.21 \\
B0531+21 & $1.26\times 10^3$  & 25 & 9.99 \\
J0537-6910 & $4.93\times 10^3$ & 23 & 53.16 \\
J0729-1448 & $3.52\times 10^4$ & 5 & 11.30 \\
B0833-45 & $1.13\times 10^4$ & 17 & 8.04 \\
B1046-58 & $2.03\times 10^4$ & 6 & 7.66 \\
J1105-6107 & $6.33\times 10^4$ & 5 & 7.50 \\
B1338-62 & $1.21\times 10^4$ & 23 & 7.18 \\
J1413-6141 & $1.36\times 10^4$ & 7 & 6.37 \\
J1420-6048 & $1.30\times 10^4$ & 5 & 6.32 \\
B1757-24 & $1.55\times 10^4$ & 5 & 3.44 \\
B1800-21 & $1.58\times 10^4$ & 5 & 5.05 \\
B1823-13 & $2.14\times 10^4$ & 6 & 4.28 \\
J1841-0524 & $3.02\times 10^4$ & 5 & 5.50 \\
B1951+32 & $1.07\times 10^5$ & 6 & 7.46 \\
J2229+6114 & $1.05\times 10^4$ & 6 & 9.31 \\
  \noalign{\smallskip}\hline
\end{tabular}
\tablenotes{1}{0.8\textwidth}{From ATNF Pulsar Catalogue~\citep{ATNF}(http://www.atnf.csiro.au/people/pulsar/psrcat/)}

\ \

\tablenotes{2}{0.8\textwidth}{From Jodrell Bank glitch-tables~\citep{Espinoza2011}(http://www.jb.man.ac.uk/pulsar/glitches/html)}

\end{table}

Now we make some discussions about why some pulsars are out of our expectation, i.e. have no glitches satisfying $\Delta\Omega/\Omega<\Delta\Omega/\Omega(E_{\rm re}=10^{35}\rm erg)$.
Comparing the data in Table~\ref{Table 1}, we can see that the spin-down ages of pulsars in the lower part of Table~\ref{Table 1}  are generally smaller than that in the upper part.
As demonstrated before, we identify the SN-triggered glitches as the ones whose sizes satisfy $\Delta\Omega/\Omega<\Delta\Omega/\Omega(E_{\rm re}=10^{35}\rm erg)$, which means that we assume that all of the pulsars have the same amount of released energy during normal glitches.
This assumption is certainly over simplified, and different pulsars should have different values of $E_{\rm re}$.
The values of $E_{\rm re}$ of young pulsars would be larger than that of old pulsars, because the former ones are more active.
The pulsars in the lower part of Table~\ref{Table 1} might mean that we should impose some larger values of $E_{\rm re}$ for the normal glitches.
However, at present stage we are lack of a reliable description about the difference of $E_{\rm re}$ between different pulsars, so we only use the criterion $\Delta\Omega/\Omega<\Delta\Omega/\Omega(E_{\rm re}=10^{35}\rm erg)$ to identify the SN-triggered glitches.
The result shown in Figure 1 is only the first attempt to reveal the  correlation between the glitch behaviors and the location of pulsars inside the Milky Way.
Whether the collision between SNs and pulsars could be a glitch trigger needs to be tested by further observations.

\section{Conclusions and discussions}

In this paper we propose a probability that small glitches of pulsars could be the result of accreting SNs which are relics of QCD phase transition in the early Universe.
This kind of glitch-trigger mechanism could be a possible astrophysical consequence of SNs that formed in the early Universe and survived to exist in our present Galaxy.
The trigger mechanism is demonstrated in the framework of star-quake of solid strange stars.
Under the typical parameters in solid strange star model,  the collision between an SN and a pulsar could lead to the broken of the solid body (or crust) of the star, which would give rise to a glitch.
The amount of energy releasing of such SN-triggered glitches would be smaller than that of the normal glitches, as the latter occur when the elastic energy accumulated reaches a critical value.
In the star-quake model the glitch size $\Delta\Omega/\Omega$ is proportional to the amount of energy releasing $E_{\rm re}$, so the sizes of SN-triggered glitches should be smaller than that of the normal glitches.
Taking into account that some uncertainties of both $E_{\rm re}$ and the mass $M$ for different pulsars could be mutually eliminated to some degree, we distinguish the SN-triggered glitches from the normal glitches by their sizes with $\Delta\Omega/\Omega<\Delta\Omega/\Omega(E_{\rm re}=10^{35}\rm erg)$, under the typical values for the mass $M$ and radius $R$.

Combined with the assumption that a huge amount of dark matter (we assume 10\% in this paper) is in the form of SNs, and the data from the glitch data in ATNF Pulsar Catalogue and Jodrell Bank glitch table, we compare the collision rates of pulsars and SNs (i.e. the expect rates of glitches with $\Delta\Omega/\Omega<\Delta\Omega/\Omega(E_{\rm re}=10^{35}\rm erg)$) and the observational data.
Among 25 pulsars that we choose whose recorded numbers of glitches are larger than 5, 9 pulsars have glitch events with $\Delta\Omega/\Omega<\Delta\Omega/\Omega(E_{\rm re}=10^{35}\rm erg)$ whereas 16 pulsars have no such event.
The incidences of glitches with $\Delta\Omega/\Omega<\Delta\Omega/\Omega(E_{\rm re}=10^{35}\rm erg)$ of the 9 pulsars  exhibit the tendency consistent with the collision rates of pulsars and SNs, if the baryon number of each SN $A\sim 10^{33}$.
The remaining 16 pulsars are generally younger than the 9 pulsars, so the zero event rate might result from the fact that younger pulsars should release more energy during glitches.

The mass (or baryon number $A$) of each SN is unknown, which depends on the properties of hot quark-gluon plasma and the state of dense matter.
From astrophysical point of view, we want to give some constraints combined with observations.
It seems that the constraints from Big Bang Nucleosynthesis ($A\gtrsim 10^{25}$), from the supermassive black holes at high redshifts ($A\lesssim 10^{35}$) and in this paper from pulsar glitches ($A\sim 10^{33}$) are consistent.
Certainly, SNs should have some mass spectrum instead of a uniform mass, so the mass corresponding to the baryon number we give above could be seem as the peak value of a very steep spectrum.
More detailed studies about the spectrum are expected in the future.

The result we give in this paper is the first attempt to show the possibility that SNs could be a kind of glitch-trigger, demonstrated as the fact that there are some correlations between the glitch behaviors and the location of pulsars inside the Milky Way.
The crude assumptions we make in this paper should be improved in further work, e.g. some more quantitative descriptions about the dependences of $E_{\rm re}$ on pulsar masses and ages.
In addition, a detailed description about the mechanism of small glitches as the result of collisions between SNs and pulsars also depends on the states of matter inside both pulsars and SNs.
Anyhow, whether the SNs could be a kind of glitch trigger remains to be explored.

The glitches with small sizes are now difficult for us to detect, so in some senses, to demonstrate the relation between SNs and glitches by the comparison between expected and observed events from recent data should be inadequate.
However, we still try to do this because some positive results have been shown, which makes it possible that more data in the future would give us a clear answer.
A more convincing demonstration about whether SNs could trigger small glitches will be done if only when more small glitches would be detected.
This work emphasizes the importance of detecting small glitches, which is also the task of SKA (the Square Kilometre Array) that has exquisite timing precision and will dramatically increase the number of sources in surveys~\citep{SKA}.
We are then expecting to test our model by future advance radio facilities, e.g., SKA and Chinese FAST (Five-hundred-meter Aperture Spherical Telescope).

\normalem
\begin{acknowledgements}
We would like to thank useful discussions at the research groups in Xinjiang University, Peking University and Xinjiang Astronomical Observatory, and an anonymous referee for helpful comments.
This work is supported by the National Natural Science Foundation of China (11203018), the West Light Foundation (XBBS-2014-23), the Science Project of Universities in Xinjiang (XJEDU2012S02) and the Doctoral Science Foundation of Xinjiang University (BS120107).
\end{acknowledgements}

\label{lastpage}

\end{document}